\documentclass[11pt]{article}
\usepackage{amssymb,amsmath,amsfonts}
\usepackage{graphicx}
\usepackage{graphics}
\usepackage{eepic,epsfig}

\begin{document}

\title{ COHERENT PAIR PRODUCTION IN DEFORMED CRYSTALS WITH A COMPLEX BASE}
\author{A. R. MKRTCHYAN, A. A. SAHARIAN\thanks{%
E-mail: saharyan@server.physdep.r.am} and V. V. PARAZIAN \\
\textit{\small Institute of Applied Problems in Physics, 25
Nersessian Str., 375014 Yerevan, Armenia}} \maketitle

\begin{abstract}
We investigate the coherent electron-positron pair creation by high-energy
photons in a periodically deformed single crystal with a complex base. The
formula for the corresponding differential cross-section is derived for an
arbitrary deformation field. The conditions are specified under which the
influence of the deformation is considerable. The case is considered in
detail when the photon enters into the crystal at small angles with respect
to a crystallographic axis. The results of the numerical calculations are
presented for $\mathrm{SiO}_{2}$ single crystal and Moliere parametrization
of the screened atomic potentials in the case of the deformation field
generated by the acoustic wave of $S$ type. In dependence of the parameters,
the presence of deformation can either enhance or reduce the pair creation
cross-section. This can be used to control the parameters of the positron
sources for storage rings and colliders.
\end{abstract}

\bigskip


\bigskip


\bigskip

\section{Introduction}

\label{sec1}

The investigation of high-energy electromagnetic processes in crystals is of
interest not only from the viewpoint of underlying physics but also from the
viewpoint of practical applications. If the formation length exceeds the
interatomic spacing, the interference effects from all atoms within this
length are important and the cross-sections of the electromagnetic processes
in crystals can change essentially compared with the corresponding
quantities for a single atom (see, for instance, \cite{TerMik}-\cite{Baie89}
and references therein). From the point of view of controlling the
parameters of various processes in a medium, it is of interest to
investigate the influence of external fields, such as acoustic waves,
temperature gradient etc., on the corresponding characteristics. The
considerations of concrete processes, such as diffraction radiation \cite%
{MkrtDR}, transition radiation \cite{Grigtrans}, parametric X-radiation \cite%
{Mkrt91}, channelling radiation \cite{Mkrtch1}, bremsstrahlung by
high-energy electrons \cite{Mkrtbrem}, have shown that the external fields
can essentially change the angular-frequency characteristics of the
radiation intensities. Motivated by the fact that the basic source for the
creation of positrons for high-energy electron-positron colliders is the
electron-positron pair creation by high-energy photons, in \cite{Mkrt03} we
have investigated the influence of the hypersonic wave excited in a crystal
on this process. The case of simplest crystal with one atom in the lattice
base and the sinusoidal deformation field generated by the hypersound were
considered. To have an essential influence of the acoustic wave
high-frequency hypersound is needed. Usually this type of waves is excited
by high-frequency electromagnetic field through the piezoelectric effect in
crystals with a complex base. In the present paper we generalize the results
of \cite{Mkrt03} for crystals with a complex base and for acoustic waves
with an arbitrary profile. The numerical calculations are carried out for
the quartz single crystal and for the photons of energy 100 GeV. The results
of the numerical calculations on the base of the formulae given in this
paper for the pair creation cross-section by the photons of energy 3.5 GeV
are presented in our recent paper \cite{Mkrt05}, where the scheme of
experimental setup is proposed for the corresponding measurements on the
photon channel of the Yerevan synchrotron.

The paper is organized as follows. In the next section we derive the general
formula for the coherent part of the pair creation cross-section averaged on
thermal fluctuations and the conditions are specified under which the
influence of the deformation field can be considerable. The analysis of the
general formula in the cases when the photon enters into the crystal at
small angles with respect to crystallographic axes or planes is given in
Sec. \ref{sec3}. The results of the numerical calculations for the
cross-section as a function of the positron energy and the amplitude of the
external excitation are presented. Sec. \ref{sec4} summarizes the main
results of the paper.

\section{Cross-section for the coherent pair creation}

\label{sec2}

Consider the creation of electron-positron pairs by high-energy photons in a
crystal. We denote by $(\omega ,\mathbf{k})$, $(E_{+},\mathbf{p}_{+})$, and $%
(E_{-},\mathbf{p}_{-})$ the energies and momenta for the photon, positron,
and electron respectively. In the discussion below the collective index $n$
enumerates the elementary cell and the superscript $j$ enumerates the atoms
in a given cell of a crystal. Let $d^{4}\sigma _{0}/dE_{+}d^{3}q=\left\vert
u_{\mathbf{q}}^{(j)}\right\vert ^{2}\sigma _{0}(\mathbf{q})$ be the
cross-section for the electron-positron pair creation on an individual $j$%
-th atom as a function of momentum transfer $\mathbf{q}=\mathbf{k}-\mathbf{p}%
_{+}-\mathbf{p}_{-}$ and $u_{\mathbf{q}}^{(j)}$ is the Fourier transform of
the potential for the atom. Usually one writes the quantity $u_{\mathbf{q}%
}^{(j)}$ in the form $4\pi Z_{j}e^{2}\left[ 1-F^{(j)}(q)\right] /q^{2}$,
where $Z_{j}$ and $F^{(j)}(q)$ are the number of electrons and the atomic
form-factor for the $j$-th atom. In the numerical calculations below we will
use the Moliere parametrization of the screened atomic potential. The
differential cross-section for the pair creation in a crystal can be written
in the form (see, for example, \cite{TerMik,Shulga})%
\begin{equation}
\sigma (\mathbf{q})\equiv \frac{d^{4}\sigma }{dE_{+}d^{3}q}=\left\vert
\sum_{n,j}u_{\mathbf{q}}^{(j)}e^{i\mathbf{qr}_{n}^{(j)}}\right\vert
^{2}\sigma _{0}(\mathbf{q}),  \label{sigma1}
\end{equation}%
where $\mathbf{r}_{n}^{(j)}$ is the position of an atom in the crystal. The
interference factor in Eq. (\ref{sigma1}) is responsible for coherent
effects arising due to periodical arrangement of the atoms in a crystal. At
nonzero temperature one has $\mathbf{r}_{n}^{(j)}=\mathbf{r}_{n0}^{(j)}+%
\mathbf{u}_{tn}^{(j)}$, where $\mathbf{u}_{tn}^{(j)}$ is the displacement of
$j$-th atom with respect to the equilibrium positions $\mathbf{r}_{n0}^{(j)}$
due to the thermal vibrations. After averaging on thermal fluctuations, the
cross-section is written in the form (see, for instance, \cite{TerMik} for
the case of a crystal with a simple cell)%
\begin{equation}
\sigma (\mathbf{q})=\left\{ N\sum_{j}\left\vert u_{\mathbf{q}%
}^{(j)}\right\vert ^{2}\left( 1-e^{-q^{2}\overline{u_{t}^{(j)2}}}\right)
+\left\vert \sum_{n,j}u_{\mathbf{q}}^{(j)}e^{i\mathbf{qr}_{n0}^{(j)}}e^{-%
\frac{1}{2}q^{2}\overline{u_{t}^{(j)2}}}\right\vert ^{2}\right\} \sigma _{0}(%
\mathbf{q}),  \label{sigma2}
\end{equation}%
where $N$ is the number of cells, $\overline{u_{t}^{(j)2}}$ is the
temperature dependent mean-squared amplitude of the thermal vibrations of
the $j$-th atom, $e^{-q^{2}\overline{u_{t}^{(j)2}}}$ is the corresponding
Debye-Waller factor. In formula (\ref{sigma2}) the first term in figure
braces does not depend on the direction of the vector $\mathbf{k}$\ and
determines the contribution of incoherent effects. The contribution of
coherent effects is presented by the second term. By taking into account the
formula for the cross-section on a single atom\ (see, e.g., \cite%
{TerMik,Shulga}), in the region of transferred momenta $q\ll m_{e}$ the
corresponding part of the cross-section in a crystal can be presented in the
form (the system of units $\hbar =c=1$ is used)
\begin{equation}
\sigma _{c}=\frac{e^{2}}{(2\pi )^{3}\omega ^{2}}\frac{q_{\perp }^{2}}{%
q_{\parallel }^{2}}\left( \frac{\omega \delta }{m_{e}^{2}}-1+\frac{2\delta }{%
q_{\parallel }}-\frac{2\delta ^{2}}{q_{\parallel }^{2}}\right) \left\vert
\sum_{n,j}u_{\mathbf{q}}^{(j)}e^{i\mathbf{qr}_{n0}^{(j)}}e^{-\frac{1}{2}q^{2}%
\overline{u_{t}^{(j)2}}}\right\vert ^{2},  \label{sigmac}
\end{equation}%
where $\mathbf{q}_{\parallel }$ and $\mathbf{q}_{\perp }$ are the components
of the vector $\mathbf{q}$ parallel and perpendicular to the direction of
the photon momentum $\mathbf{k}$, $\delta =1/l_{c}$ is the minimum
longitudinal momentum transfer, and $l_{c}=2E_{+}E_{-}/(m_{e}^{2}\omega )$
is the formation length for the pair creation process.

When external influences are present (for example, in the form of acoustic
waves) the positions of atoms in the crystal can be written as $\mathbf{r}%
_{n0}^{(j)}=\mathbf{r}_{ne}^{(j)}+\mathbf{u}_{n}^{(j)}$, where $\mathbf{r}%
_{ne}^{(j)}$\ determines the equilibrium position of an atom in the
situation without the deformation, $\mathbf{u}_{n}^{(j)}$ is the
displacement of the atom caused by the external influence. We will consider
deformations with the periodical structure:%
\begin{equation}
\mathbf{u}_{n}^{(j)}=\mathbf{u}_{0}f(\mathbf{k}_{s}\mathbf{r}_{ne}^{(j)}),
\label{un}
\end{equation}%
where $\mathbf{u}_{0}$ and $\mathbf{k}_{s}$\ are the amplitude and wave
vector corresponding to the deformation field, $f(x)$ is an arbitrary
function with the period $2\pi $, $\max f(x)=1$. In the discussion below we
will assume that $f(x)\in C^{\infty }(R)$. Note that the dependence of $%
\mathbf{u}_{n}^{(j)} $\ on the time coordinate for the case of acoustic
waves we can disregard, as for particle energies we are interested in, the
characteristic time for the change of the deformation field is much greater
compared with the passage time of particles through the crystal. For the
deformation field given by Eq. (\ref{un}) the sum over $n$ in (\ref{sigma2})
can be transformed into the form%
\begin{equation}
\sum_{n}u_{\mathbf{q}}^{(j)}e^{i\mathbf{qr}_{n0}^{(j)}}=\sum_{m=-\infty
}^{\infty }F_{m}(\mathbf{qu}_{0})\sum_{n}u_{\mathbf{q}}^{(j)}e^{i\mathbf{q}%
_{m}\mathbf{r}_{ne}^{(j)}},  \label{sum1}
\end{equation}%
where $\mathbf{q}_{m}=\mathbf{q}+m\mathbf{k}_{s}$ and $F_{m}(x)$\ is the
Fourier-transform of the function $e^{ixf(t)}$:
\begin{equation}
F_{m}(x)=\frac{1}{2\pi }\int_{-\pi }^{\pi }e^{ixf(t)-imt}dt.  \label{Fm}
\end{equation}%
Below we need to have the asymptotic behavior of this function for large
values $m$. For a fixed $x$ and under the assumptions for the function $f(x)$
given above, by making use the stationary phase method we can see that $%
F_{m}(x)\sim \mathcal{O}(|m|^{-\infty })$ for $m\rightarrow \infty $.

For a lattice with a complex cell the coordinates of the atoms can be
presented in the form $\mathbf{r}_{ne}^{(j)}=\mathbf{R}_{n}+\mathbf{\rho }%
^{(j)}$, where $\mathbf{R}_{n}$\ determines the positions of the atoms for
one of primitive lattices, and $\mathbf{\rho }^{(j)}$\ are the equilibrium
positions for other atoms inside $n$-th elementary cell with respect to $%
\mathbf{R}_{n}$. \ By taking into account this, one obtains%
\begin{equation}
\sum_{m=-\infty }^{\infty }F_{m}(\mathbf{qu}_{0})\sum_{j,n}u_{\mathbf{q}%
}^{(j)}e^{-\frac{1}{2}q^{2}\overline{u_{t}^{(j)2}}}e^{i\mathbf{q}_{m}\mathbf{%
r}_{ne}^{(j)}}=\sum_{m=-\infty }^{\infty }F_{m}(\mathbf{qu}_{0})S(\mathbf{q},%
\mathbf{q}_{m})\sum_{n}e^{i\mathbf{q}_{m}\mathbf{R}_{n}},  \label{sum2}
\end{equation}%
where%
\begin{equation}
S(\mathbf{q},\mathbf{q}_{m})=\sum_{j}u_{\mathbf{q}}^{(j)}e^{i\mathbf{q}_{m}%
\mathbf{\rho }^{(j)}}e^{-\frac{1}{2}q^{2}\overline{u_{t}^{(j)2}}}
\label{Formfactor1}
\end{equation}%
is the factor determined by the structure of the elementary cell. For thick
crystals the sum over cells in (\ref{sum2}) can be presented as a sum over
the reciprocal lattice:%
\begin{equation}
\sum_{n}e^{i\mathbf{q}_{m}\mathbf{R}_{n}}=\frac{(2\pi )^{3}}{\Delta }\sum_{%
\mathbf{g}}\delta \left( \mathbf{q}-\mathbf{g}_{m}\right) ,\;\mathbf{g}_{m}=%
\mathbf{g}-m\mathbf{k}_{s}  \label{sum4}
\end{equation}%
where $\Delta $ is the unit cell volume, and $\mathbf{g}$\ is the reciprocal
lattice vector. Due to the $\delta $-function in this formula, the
corresponding momentum conservation is written in the form%
\begin{equation}
\mathbf{k}=\mathbf{p}_{+}+\mathbf{p}_{-}+\mathbf{g}-m\mathbf{k}_{s},
\label{momcons}
\end{equation}%
where $-m\mathbf{k}_{s}$ stands for the momentum transfer to the external
field. As the main contribution into the coherent part of the cross-section
comes from the longitudinal momentum transfer of an order $\delta $, the
influence of the external excitation may be considerable if $|m|k_{s}$ is of
an order $\delta $. The corresponding condition will be specified below.
Another consequence of the $\delta $-function in (\ref{sum4}) is that the
function (\ref{Fm}) enters into the cross-section in the form $F_{m}(\mathbf{%
g}_{m}\mathbf{u}_{0})$. Now it can be seen that in the sum over $m$ in (\ref%
{sum2}) the main contribution comes from the terms for which $\left\vert m%
\mathbf{k}_{s}\mathbf{u}_{0}\right\vert \lesssim \left\vert \mathbf{gu}%
_{0}\right\vert $, or equivalently $\left\vert m\right\vert \lesssim \lambda
_{s}/a$, where $\lambda _{s}=2\pi /k_{s}$ is the wavelength of the external
excitation, and $a$\ is of the order of the lattice spacing. Indeed, for the
terms with $\left\vert m\mathbf{k}_{s}\mathbf{u}_{0}\right\vert \gg
\left\vert \mathbf{gu}_{0}\right\vert $ one has $F_{m}(\mathbf{g}_{m}\mathbf{%
u}_{0})\approx F_{m}(m\mathbf{k}_{s}\mathbf{u}_{0})$, and the phase of the
integrand in (\ref{Fm}) is equal to $m\left[ \mathbf{k}_{s}\mathbf{u}%
_{0}f(t)-t\right] $. Under the condition $\left\vert \mathbf{k}_{s}\mathbf{u}%
_{0}f^{\prime }(t)\right\vert <1$ this phase has no stationary point and one
has $F_{m}(m\mathbf{k}_{s}\mathbf{u}_{0})=\mathcal{O}(|m|^{-\infty })$, $%
m\rightarrow \infty $ and the corresponding contribution is strongly
suppressed. By taking into account that for practically important cases one
has $\mathbf{k}_{s}\mathbf{u}_{0}\sim u_{0}/\lambda _{s}\ll 1$, we see that
the assumption made means that the derivative $f^{\prime }(t)$ is not too
large. By making use of the formulae given above, the square of the modulus
for the sum (\ref{sum1}) takes the form%
\begin{eqnarray}
\left\vert \sum_{j,n}u_{\mathbf{q}}^{(j)}e^{i\mathbf{qr}_{n0}}e^{-\frac{1}{2}%
q^{2}\overline{u_{t}^{(j)2}}}\right\vert ^{2} &=&\sum_{m,\mathbf{g}}F_{m}(%
\mathbf{g}_{m}\mathbf{u}_{0})S(\mathbf{g}_{m},\mathbf{g})\delta \left(
\mathbf{q}-\mathbf{g}_{m}\right)   \notag \\
&&\times \sum_{m^{\prime }}F_{m^{\prime }}^{\ast }(\mathbf{qu}_{0})S^{\ast }(%
\mathbf{q},\mathbf{q}_{m^{\prime }})\sum_{n}e^{i(m-m^{\prime })\mathbf{k}_{s}%
\mathbf{R}_{n}}.  \label{sum3}
\end{eqnarray}%
In the case $m^{\prime }\neq m$ the main contribution into the sum over $n$
comes from the summands satisfying the condition $(m^{\prime }-m)\mathbf{k}%
_{s}=\mathbf{g}$. It follows from here that $\left\vert m^{\prime
}-m\right\vert \gtrsim \lambda _{s}/a$. By taking into account that the main
contribution into the sum over $m$ comes from $\left\vert m\right\vert
\lesssim \lambda _{s}/a$, we conclude that $\left\vert m^{\prime
}\right\vert \gtrsim \lambda _{s}/a\gg 1$. Now we see that for the function $%
F_{m^{\prime }}(\mathbf{qu}_{0})$ the ratio of the order to the argument is
estimated as $m^{\prime }/\mathbf{qu}_{0}\sim \lambda _{s}/2\pi u_{0}$.
Combining this with the asymptotic behavior of the function $F_{m}(x)$ for
large values of the order given in the paragraph after formula (\ref{Fm}),
we see that under the condition $u_{0}/\lambda _{s}\ll 1$\ the contribution
of the terms with $m\neq m^{\prime }$\ in the sum (\ref{sum3}) is small
compared to the diagonal terms. In the case $m=m^{\prime }$\ the sum over $n$
in the left hand side is equal to the number of cells, $N$, in a crystal and
the square of the modulus for the sum on the left of Eq. (\ref{sum3}) can be
written as%
\begin{equation}
\left\vert \sum_{j,n}u_{\mathbf{q}}^{(j)}e^{i\mathbf{qr}_{n0}}e^{-\frac{1}{2}%
q^{2}\overline{u_{t}^{(j)2}}}\right\vert ^{2}=N\frac{(2\pi )^{3}}{\Delta }%
\sum_{m,\mathbf{g}}\left\vert F_{m}(\mathbf{g}_{m}\mathbf{u}_{0})\right\vert
^{2}\left\vert S(\mathbf{g}_{m},\mathbf{g})\right\vert ^{2}.  \label{sum6}
\end{equation}%
Substituting this expression into formula (\ref{sigmac}) and integrating
over the vector $\mathbf{q}$ by using the $\delta $-function, for the
cross-section one obtains%
\begin{equation}
d\sigma =\int \sigma (\mathbf{q})d^{3}q=N_{0}(d\sigma _{n}+d\sigma _{c}),
\label{dsigman+c}
\end{equation}%
with $d\sigma _{n}$ and $d\sigma _{c}$ being the incoherent and coherent
parts of the cross-section per atom and $N_{0}$ is the number of atoms in
the crystal. The coherent part of the cross-section is determined by the
formula%
\begin{equation}
\frac{d\sigma _{c}}{dE_{+}}=\frac{e^{2}N}{\omega ^{2}N_{0}\Delta }\sum_{m,%
\mathbf{g}}\frac{g_{m\perp }^{2}}{g_{m\parallel }^{2}}\left( \frac{\omega
^{2}}{2E_{+}E_{-}}-1+\frac{2\delta }{g_{m\parallel }}-\frac{2\delta ^{2}}{%
g_{m\parallel }^{2}}\right) \left\vert F_{m}(\mathbf{g}_{m}\mathbf{u}%
_{0})\right\vert ^{2}\left\vert S(\mathbf{g}_{m},\mathbf{g})\right\vert ^{2},
\label{dsigmac/dE+}
\end{equation}%
where the vector $\mathbf{g}_{m}$ is defined by relation (\ref{sum4}) and
the summation goes under the constraint $g_{m\parallel }\geq \delta $. For a
simplest crystal with one atom in the lattice base and sinusoidal
deformation field, $f(z)=\sin (z+\varphi _{0})$, the formula (\ref%
{dsigmac/dE+}) reduces to the result obtained in \cite{Mkrt03}. Note that
for this type of deformation one has the Fourier-transform%
\begin{equation}
F_{m}(x)=e^{im\varphi _{0}}J_{m}(x),  \label{Fmzsin}
\end{equation}%
with the Bessel function $J_{m}(x)$.

The formula for the pair creation in an undeformed crystal is obtained from (%
\ref{dsigmac/dE+}) taking $\mathbf{u}_{0}=0$. In this limit the contribution
of the term with $m=0$ remains only with $F_{0}(0)=1$. The corresponding
formula for a crystal with simple lattice base can be found, for instance,
in \cite{TerMik,Shulga}. Now we see that formula (\ref{dsigmac/dE+}) differs
from the formula in an undeformed crystal by the replacement $\mathbf{g}%
\rightarrow \mathbf{g}_{m}$, and by an additional summation over $m$ with
weights $\left\vert F_{m}(\mathbf{g}_{m}\mathbf{u}_{0})\right\vert ^{2}$.
This corresponds to the presence of an additional one-dimensional
superlattice with the period $\lambda _{s}$ and the reciprocal lattice
vector $m\mathbf{k}_{s}$, $m=0,\pm 1,\pm 2,\ldots $. As the main
contribution into the cross-section comes from the terms with $g_{m\parallel
}\sim \delta $, the influence of the deformation field may be considerable
if $|mk_{s\parallel }|\gtrsim \delta $. Combining this with the previous
estimates we find the following condition: $u_{0}/\lambda _{s}\gtrsim a/4\pi
^{2}l_{c}$. At high energies one has $a/l_{c}\ll 1$ and this condition can
be consistent with the condition $u_{0}/\lambda _{s}\ll 1$.

\section{Limiting cases and numerical results}

\label{sec3}

If the photon moves in a non-oriented crystal, in formula (\ref{dsigmac/dE+}%
) the summation over $\mathbf{g}$\ can be replaced by the integration and
the pair creation cross-section coincides with that in an amorphous medium.
Note that for an amorphous medium the ratio $u_{0}/\lambda _{s}$ can be
relatively larger compared with the corresponding quantity in crystals.
Below we consider the case when the photon enters into the crystal at small
angle $\theta $ with respect to the crystallographic $z$-axis of the
orthogonal lattice. The corresponding reciprocal lattice vector components
are $g_{i}=2\pi n_{i}/a_{i}$, $n_{i}=0,\pm 1,\pm 2,\ldots $, where $a_{i}$, $%
i=1,2,3$, are lattice constants in the corresponding directions. We can write%
\begin{equation}
g_{m\parallel }=g_{mz}\cos \theta +(g_{mx}\sin \alpha +g_{my}\cos \alpha
)\sin \theta ,  \label{gmpar}
\end{equation}%
where $\alpha $ is the angle between the projection of the vector $\mathbf{k}
$ on the plane $(x,y)$ and $y$-axis. For small angles $\theta $ the main
contribution into the cross-section comes from the summands with $g_{z}=0$
and from formula (\ref{dsigmac/dE+}) one finds%
\begin{equation}
\frac{d\sigma _{c}}{dE_{+}}\approx \frac{e^{2}N}{\omega ^{2}N_{0}\Delta }%
\sum_{m,g_{x},g_{y}}\frac{g_{\perp }^{2}}{g_{m\parallel }^{2}}\left( \frac{%
\omega ^{2}}{2E_{+}E_{-}}-1+\frac{2\delta }{g_{m\parallel }}-\frac{2\delta
^{2}}{g_{m\parallel }^{2}}\right) \left\vert F_{m}(\mathbf{g}_{m}\mathbf{u}%
_{0})\right\vert ^{2}\left\vert S(\mathbf{g}_{m},\mathbf{g})\right\vert ^{2},
\label{dsigmac1/dE+}
\end{equation}%
where $g_{\perp }^{2}=g_{x}^{2}+g_{y}^{2}$, and the summation goes over the
region $g_{m\parallel }\geq \delta $ with%
\begin{equation}
g_{m\parallel }\approx -mk_{z}+(g_{x}\sin \alpha +g_{y}\cos \alpha )\theta .
\label{gmpar1}
\end{equation}%
Note that in the arguments of the functions $F_{m}$ and $S$ we have $\mathbf{%
g}_{m}\approx (g_{x},g_{y},0)$.

If the photon moves far from the crystallographic planes (angles $\alpha $
and $\pi /2-\alpha $ are not small compared with unity), in Eq. (\ref%
{dsigmac1/dE+}) the summation over $g_{x}$ and $g_{y}$ can be replaced by
the integration, $\sum_{g_{x},g_{y}}\rightarrow (a_{1}a_{2}/4\pi ^{2})\int
dg_{x}dg_{y}$, and one receives%
\begin{eqnarray}
\frac{d\sigma _{c}}{dE_{+}} &\approx &\frac{e^{2}N\omega ^{-2}}{4\pi
^{2}N_{0}a_{3}}\sum_{m}\int dg_{x}dg_{y}\frac{g_{\perp }^{2}}{g_{m\parallel
}^{2}}\left( \frac{\omega ^{2}}{2E_{+}E_{-}}-1+\frac{2\delta }{g_{m\parallel
}}-\frac{2\delta ^{2}}{g_{m\parallel }^{2}}\right)  \label{dsigmac2/dE+} \\
&&\times \left\vert F_{m}(\mathbf{g}_{m}\mathbf{u}_{0})\right\vert
^{2}\left\vert S(\mathbf{g}_{m},\mathbf{g})\right\vert ^{2},  \notag
\end{eqnarray}%
where the integration goes under the constraint $g_{m\parallel }\geq \delta $%
.

We now assume that the photon enters into the crystal at small angle $\theta
$ with respect to the crystallographic axis $z$ and near the
crystallographic plane $(y,z)$ (the angle $\alpha $ is small). In this case
with the change of $\delta $, the sum over $g_{x}$ and $g_{y}$ will drop
sets of terms which leads to the abrupt change of the corresponding
cross-section. Two cases have to be distinguished. Under the condition $%
\delta \sim 2\pi \theta /a_{2}$, in Eq. (\ref{dsigmac1/dE+}) for the
longitudinal component one has%
\begin{equation}
g_{m\parallel }\approx -mk_{z}+\theta g_{y}\geq \delta .  \label{gmpar2}
\end{equation}%
In this case the summation over the component $g_{x}$ can be replaced by the
integration, $\sum_{g_{x}}\rightarrow (a_{1}/2\pi )\int dg_{x}$, and we have
the formula%
\begin{eqnarray}
\frac{d\sigma _{c}}{dE_{+}} &\approx &\frac{e^{2}N\omega ^{-2}}{2\pi
N_{0}a_{2}a_{3}}\sum_{m,g_{y}}\int dg_{x}\frac{g_{\perp }^{2}}{g_{m\parallel
}^{2}}\left( \frac{\omega ^{2}}{2E_{+}E_{-}}-1+\frac{2\delta }{g_{m\parallel
}}-\frac{2\delta ^{2}}{g_{m\parallel }^{2}}\right)  \label{dsigmac3/dE+} \\
&&\times \left\vert F_{m}(\mathbf{g}_{m}\mathbf{u}_{0})\right\vert
^{2}\left\vert S(\mathbf{g}_{m},\mathbf{g})\right\vert ^{2}.  \notag
\end{eqnarray}%
This formula can be further simplified under the assumption $\mathbf{u}%
_{0}\perp \mathbf{a}_{1}$. In this case in the argument of the function $%
F_{m}$ one has $\mathbf{g}_{m}\mathbf{u}_{0}\approx g_{y}u_{0y}$ and we
obtain the formula%
\begin{equation}
\frac{d\sigma _{c}}{dE_{+}}\approx \frac{e^{2}N\omega ^{-2}}{2\pi
N_{0}a_{2}a_{3}}\sum_{m,g_{y}}\frac{\left\vert F_{m}(g_{y}u_{0y})\right\vert
^{2}}{g_{m\parallel }^{2}}\left( \frac{\omega ^{2}}{2E_{+}E_{-}}-1+\frac{%
2\delta }{g_{m\parallel }}-\frac{2\delta ^{2}}{g_{m\parallel }^{2}}\right)
\int dg_{x}g_{\perp }^{2}\left\vert S(\mathbf{g}_{m},\mathbf{g})\right\vert
^{2},  \label{dsigmac35/dE+}
\end{equation}%
with an effective structure factor determined by the integral on the right.

In the second case we assume that $\delta \sim 2\pi \theta \alpha /a_{1}$.
Now the main contribution into the sum in Eq. (\ref{dsigmac1/dE+}) comes
from the terms with $g_{y}=0$ and two summations remain: over $m$ and over $%
n_{1}$. The formula for the cross-section takes the form%
\begin{equation}
\frac{d\sigma _{c}}{dE_{+}}\approx \frac{e^{2}N}{\omega ^{2}N_{0}\Delta }%
\sum_{m,n_{1}}\frac{g_{\perp }^{2}}{g_{m\parallel }^{2}}\left( \frac{\omega
^{2}}{2E_{+}E_{-}}-1+\frac{2\delta }{g_{m\parallel }}-\frac{2\delta ^{2}}{%
g_{m\parallel }^{2}}\right) \left\vert F_{m}(\mathbf{g}_{m}\mathbf{u}%
_{0})\right\vert ^{2}\left\vert S(\mathbf{g}_{m},\mathbf{g})\right\vert ^{2},
\label{dsigmac4/dE+}
\end{equation}%
where%
\begin{equation}
g_{m\parallel }\approx -mk_{z}+\psi g_{x},\;\psi =\alpha \theta ,
\label{gmpar3}
\end{equation}%
and the summation goes over the values $m$ and $n_{1}$ satisfying the
condition $g_{m\parallel }\geq \delta $.

We have carried out numerical calculations for the pair creation
cross-section for various values of parameters in the case of $\mathrm{SiO}%
_{2}$ single crystal at zero temperature. To deal with an orthogonal
lattice, we choose as an elementary cell the cell including 6 atoms of
silicon and 12 atoms of oxygen (Shrauf elementary cell \cite{Dana62}). For
this choice the $y$ and $z$ axes of the orthogonal coordinate system $(x,y,z)
$ coincide with the standard $Y$ and $Z$ axes of the quartz crystal, whereas
the angle between the axes $x$ and $X$ is equal to $\pi /6$. For the
potentials of atoms we take Moliere parametrization%
\begin{equation}
u_{\mathbf{q}}^{(j)}=\sum_{i=1}^{3}\frac{4\pi Z_{j}e^{2}\alpha _{i}}{%
q^{2}+(\chi _{i}/R_{j})^{2}},  \label{Moliere}
\end{equation}%
where $\alpha _{i}=\left\{ 0.1,0.55,0.35\right\} $, $\chi _{i}=\left\{
6.0,1.2,0.3\right\} $, and $R_{j}$ is the screening radius for the $j$-th
atom in the elementary cell. The calculations are carried out for the
sinusoidal transversal acoustic wave of the $S$ type (the corresponding
parameters can be found in Ref. \cite{Shas}) for which the vector of the
amplitude of the displacement is directed along $X$-direction of quartz
single crystal, $\mathbf{u}_{0}=(u_{0},0,0)$, and the velocity is $%
4.687\cdot 10^{5}$ cm/sec. The vector determining the direction of the
hypersound propagation lies in the plane $YZ$ and has the angle with the
axis $Z$ equal to $0.295$ rad. As the axis $z$ we choose the axis $Z$ of the
quartz crystal. The corresponding function $F(x)$ is determined by formula (%
\ref{Fmzsin}). Note that in the case of potential (\ref{Moliere}) for the
integral in formula (\ref{dsigmac35/dE+}) one has%
\begin{equation}
\int dg_{x}g_{\perp }^{2}\left\vert S(\mathbf{g}_{m},\mathbf{g})\right\vert
^{2}\approx 32\pi ^{3}e^{4}\sum_{j,j^{\prime },i,i^{\prime }}\cos \left(
g_{y}\rho _{y}^{(jj^{\prime })}\right) \frac{e^{-\rho _{x}^{(jj^{\prime })}%
\sqrt{g_{y}^{2}+\chi _{i}^{2}/R_{j}^{2}}}}{\sqrt{g_{y}^{2}+\chi
_{i}^{2}/R_{j}^{2}}}\frac{\alpha _{i}\alpha _{i^{\prime }}Z_{j}Z_{j^{\prime
}}\chi _{i}^{2}/R_{j}^{2}}{\chi _{i}^{2}/R_{j}^{2}-\chi _{i^{\prime
}}^{2}/R_{j^{\prime }}^{2}},  \label{Smol}
\end{equation}%
where we use the notation $\rho _{q}^{(jj^{\prime })}=\left\vert \rho
_{q}^{(j)}-\rho _{q}^{(j^{\prime })}\right\vert $, $q=x,y$. For the case $%
\chi _{i}/R_{j}=\chi _{i^{\prime }}/R_{j^{\prime }}$ in (\ref{Smol}) one has
uncertainty 0/0 which has to be evaluated by the Lopitale rule.

\begin{figure}[tbph]
\begin{center}
\begin{tabular}{ccc}
\epsfig{figure=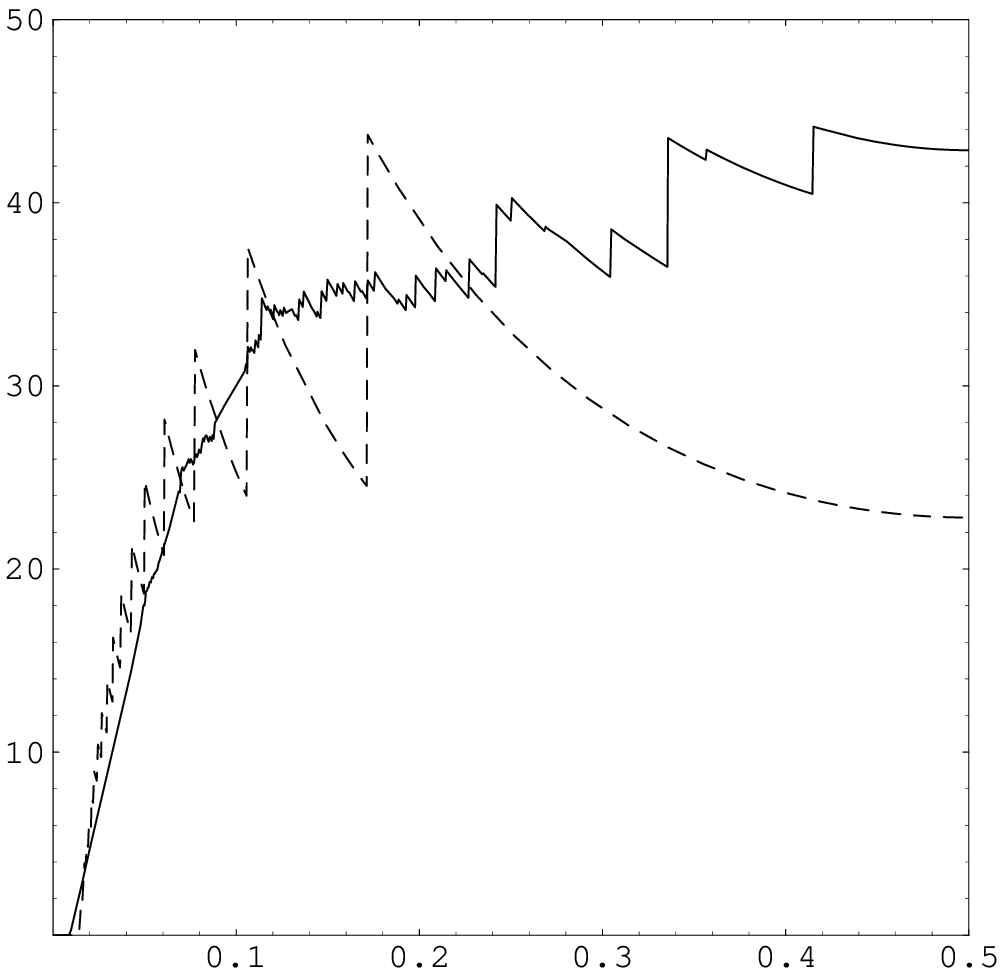,width=6cm,height=6cm} & \hspace*{0.5cm} & %
\epsfig{figure=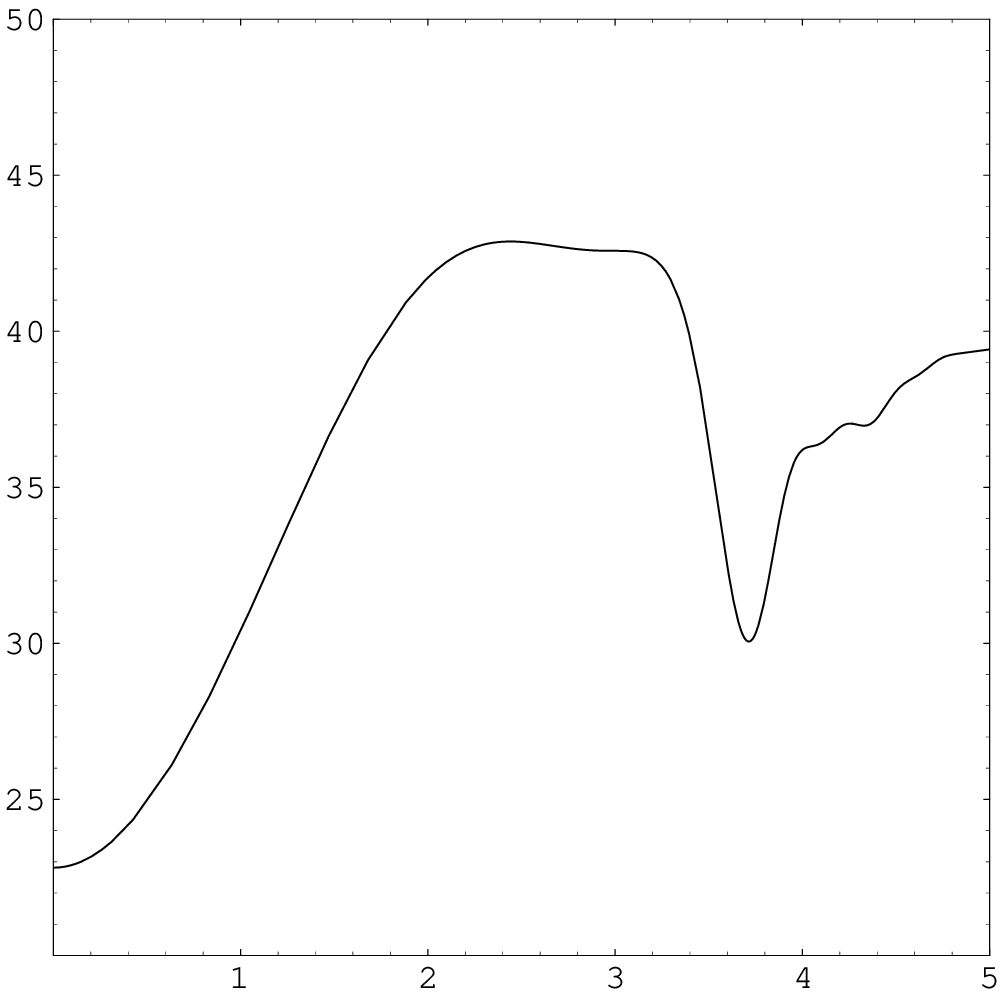,width=6cm,height=6cm}%
\end{tabular}%
\end{center}
\caption{Coherent part of the pair creation cross-section, $10^{-3}(m_{e}^{2}%
\protect\omega /e^{6})d\protect\sigma _{c}/dE_{+}$, evaluated by
formula (\protect\ref{dsigmac35/dE+}) with ( \protect\ref{Smol}),
in the quartz single crystal for the sinusoidal transversal
acoustic wave of the $S$ type with frequency 5 GHz as a function
of $E_{+}/\protect\omega $ (left panel) for $u_{0}=0$ (dashed
curve), $2\protect\pi u_{0}/a_{2}=2.4$ (full curve) and as a
function of $2\protect\pi u_{0}/a_{2}$ for the positron energy
corresponding to $E_{+}/\protect\omega =0.5$. The values for the
other parameters are as follows: $\protect\theta =0.002$ rad,
$\protect\omega =100$ GeV.} \label{fig1}
\end{figure}

Numerical calculations show that, in dependence of the values for the
parameters, the external excitation can either enhance or reduce the
cross-section of the pair creation process. As an illustration of the
enhancement in figure \ref{fig1} (left panel) we have depicted the quantity $%
d\sigma _{c}/dE_{+}$ evaluated by formula (\ref{dsigmac35/dE+}) with the
effective structure factor (\ref{Smol}) as a function of the ratio $%
E_{+}/\omega $ in the case of $\mathrm{SiO}_{2}$ single crystal and Moliere
parametrization of the screened atomic potential for $u_{0}=0$ (dashed
curve) and $2\pi u_{0}/a_{2}=2.4$ (full curve). The deformation is induced
by the transversal acoustic wave of the $S$ type with frequency 5 MHz. The
values for the other parameters are as follows: $\theta =0.002$ rad, $\omega
=100$ GeV. As the cross-section is symmetric under the replacement $%
E_{+}/\omega \rightarrow 1-E_{+}/\omega $, we have plotted the
graphs for the region $0\leq E_{+}/\omega \leq 0.5$ only. In
figure \ref{fig1} (right panel) the cross-section evaluated by
formula (\ref{dsigmac35/dE+}) is presented as a function of $2\pi
u_{0}/a_{2}$ for the positron energy corresponding to
$E_{+}/\omega =0.5$. The values of the other parameters are the
same as those for the left figure. Note that for the chosen values
of
the parameters one has $\lambda _{s}\approx 9.4\times 10^{-4}$cm, whereas $%
l_{c}\approx 3.8\times 10^{-6}$cm for the energies $E_{+}=E_{-}=50$ Gev and,
hence, $\lambda _{s}\gg l_{c}$.

\begin{figure}[tbph]
\begin{center}
\begin{tabular}{ccc}
\epsfig{figure=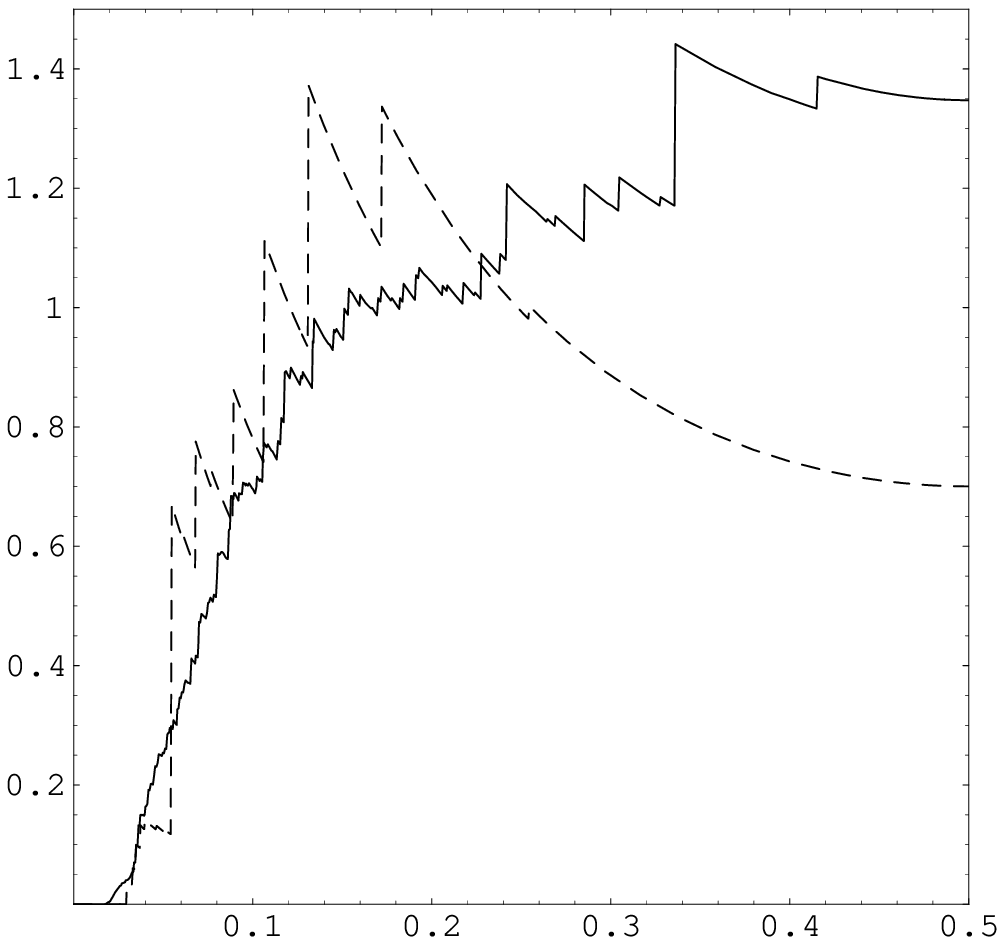,width=6cm,height=6cm} & \hspace*{0.5cm} & %
\epsfig{figure=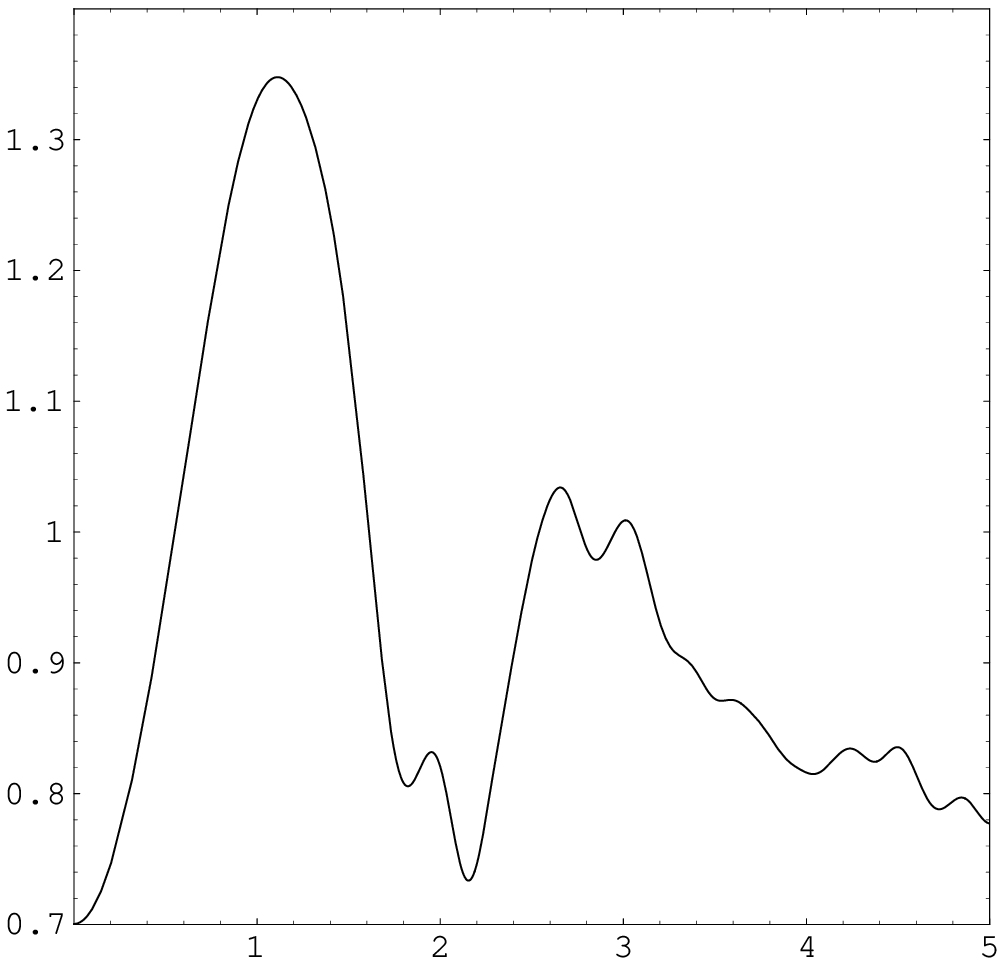,width=6cm,height=6cm}%
\end{tabular}%
\end{center}
\caption{Pair creation cross-section, $10^{-3}(m_{e}^{2}\protect\omega %
/e^{6})d\protect\sigma _{c}/dE_{+}$, evaluated by formula (\protect\ref%
{dsigmac4/dE+}), as a function of $E_{+}/\protect\omega $ (left panel) for $%
u_{0}=0$ (dashed curve), $2\protect\pi u_{0}/a_{2}=2.1$ (full curve) and as
a function of $2\protect\pi u_{0}/a_{2}$ (right panel) for the positron
energy corresponding to $E_{+}/\protect\omega =0.5$. The values for the
other parameters are as follows: $\protect\psi =0.001$ rad, $\protect\omega %
=100$ GeV.}
\label{fig2}
\end{figure}

In figure \ref{fig2} (left panel) we have presented the cross-section
evaluated by formula (\ref{dsigmac4/dE+}) as a function of the ratio $%
E_{+}/\omega $ for $u_{0}=0$ (dashed curve) and $2\pi u_{0}/a_{1}=2.1$ (full
curve) in the case $\psi =0.001$ and for the photon energy $\omega =100$
GeV. The values for the other parameters are the same as in figure \ref{fig1}%
. In figure \ref{fig2} (right panel) we have plotted the cross-section
evaluated by formula (\ref{dsigmac4/dE+}) as a function of $2\pi u_{0}/a_{1}$
for the positron energy corresponding to $E_{+}/\omega =0.5$ and for $\psi
=0.001$ rad. The values for the other parameters are the same as for the
left panel.

\section{Conclusion}

\label{sec4}

The present paper is devoted to the investigation of the electron-positron
pair creation by high-energy photons in a crystal with a complex lattice
base in the presence of deformation field of an arbitrary periodic profile.
The latter can be induced, for example, by acoustic waves. The influence of
the deformation field can serve as a possible mechanism to control the
angular-energetic characteristics of the created particles. The importance
of this is motivated by that the basic source \ to creating positrons for
high-energy colliders is the electron-positron pair creation by high-energy
photons. In a crystal the cross-section is a sum of coherent and incoherent
parts. The coherent part of the cross-section per single atom, averaged on
thermal fluctuations, is given by formula (\ref{dsigmac/dE+}). In this
formula the factor $\left\vert F_{m}(\mathbf{g}_{m}\mathbf{u}%
_{0})\right\vert ^{2}$ is determined by the function describing the
displacement of the atoms due to the deformation field, and the factor $%
\left\vert S(\mathbf{g}_{m},\mathbf{g})\right\vert ^{2}$ is determined by
the structure of the crystal elementary cell. Compared with the
cross-section in an undeformed crystal, formula (\ref{dsigmac/dE+}) contains
an additional summation over the reciprocal lattice vector $m\mathbf{k}_{s}$
of the one-dimensional superlattice induced by the deformation field. We
have argued that the influence of the deformation field on the cross-section
can be remarkable under the condition $4\pi ^{2}u_{0}/a\gtrsim \lambda
_{s}/l_{c}$. Note that for the deformation with $4\pi ^{2}u_{0}/a>1$ this
condition is less restrictive than the naively expected one $\lambda
_{s}\lesssim l_{c}$. The role of coherence effects in the pair creation
cross-section is essential when the photon enters into the crystal at small
angles with respect to a crystallographic axis. In this case the main
contribution into the coherent part of the cross-section comes from the
crystallographic planes, parallel to the chosen axis (axis $z$ in our
consideration). The behavior of this cross-section as a function on the
positron energy essentially depends on the angle $\alpha $ between the
projection of the photon momentum on the plane $(x,y)$ and $y$-axis. If the
photon moves far from the corresponding crystallographic planes, the
summation over the perpendicular components of the reciprocal lattice vector
can be replaced by the integration and the coherent part of the pair
creation cross-section is given by formula (\ref{dsigmac2/dE+}). When the
photon enters into the crystal near a crystallographic plane, two cases have
to be distinguished. For the first one $\theta \sim a_{2}/2\pi l_{c}$, the
summation over $g_{x}$ can be replaced by integration and one obtains
formula (\ref{dsigmac3/dE+}). This formula is further simplified to the form
(\ref{dsigmac35/dE+}) under the assumption $\mathbf{u}_{0}\perp \mathbf{a}%
_{1}$. In the second case one has $\psi =\alpha \theta \sim a_{1}/2\pi l_{c}$%
, and the main contribution into the cross-section comes from the
crystallographic planes parallel to the incidence plane. The corresponding
formula for the cross-section takes the form (\ref{dsigmac4/dE+}). The
numerical calculations for the cross-section are carried out in the case of $%
\mathrm{SiO}_{2}$ single crystal with the Moliere parametrization of the
screened atomic potentials and for the deformation field generated by the
transversal acoustic wave of $S$ type with frequency $5$ GHz. Examples of
numerical results are depicted in figures \ref{fig1} and \ref{fig2}. The
numerical calculations for various values of the parameters in the problem
show that, in dependence of the values for the parameters, the presence of
the deformation field can either enhance or reduce the cross-section. This
can be used to control the parameters of the positron sources for storage
rings and colliders.

\section*{Acknowledgment}

We are grateful to Levon Grigoryan and Hrant Khachatryan for valuable
discussions and suggestions. The work has been supported by Grant no. 0061
from Ministry of Education and Science of the Republic of Armenia.

\end{document}